\documentclass[12pt,a4paper]{article}

\usepackage{graphicx}

\begin{document}

\title{ Shape of Dipole Radiative Strength Function for Asymmetric Nuclei }

\author{ V.A. Plujko$^{1,2}$, M.O. Kavatsyuk$^1$, O.O. Kavatsyuk$^1$}

\date{}

\maketitle

\vspace{-1cm}
\begin{center}
$^{1}$Taras Shevchenko National University, Glushkova Str. 6,
      Kiev, Ukraine\\
$^{2}$Institute for Nuclear Research, Prosp. Nauki 47,
      Kiev, Ukraine\\
E-mail: plujko@univ.kiev.ua
\end{center}
\vspace{0.1cm}

\begin{center}
\parbox{0.8\textwidth}{\small The semiclassical method for description of the
radiative strength function is used for asymmetric
nuclei with $N\ne Z$. The theory is based on the linearized
Vlasov-Landau equations in two-component finite Fermi liquid.
The dependence of the shape $E1$ strength on the coupling
constant between proton and neutron subsystems was studied.}
\end{center}

\section{Introduction}

At the present time the properties of nuclei far from
$\beta$-stability valley are widely investigated \cite{hamamoto,abros}.
In this case the experimental data and theoretical predictions
for photo-absorption and $\gamma$-emission processes
are needed for the astrophysics and some applied problems. These processes
can be calculated by the use of radiative strength (RS) functions.
The RS function is determined by response function heated nuclei on
external electromagnetic field. The semiclassical approach \cite{Land1,torro}
gives the simple way to calculate the response function. It includes both
self-consistence mean field and residual interaction.
In this contribution the semiclassical Fermi-liquid
model \cite{Land1,torro}
is generalized to calculation response function in asymmetric nuclei with
taking into account retardation effects \cite{kolom,acta} in collision
integral. Note that the two-component Fermi-liquid system was
investigated in the Ref. \cite{china}. But this model doesn't include
interaction $u^{\alpha\alpha'}$ between nucleon subsystems as well
as difference between Fermi-energies of neutrons and protons.
We avoid these simplification. The self-consistent mean field was
taken with the use of separable interaction potential.

The nucleus is considered as a drop of two-component Fermi-liquid
with phase-space distribution function
$f^{(\alpha)}(\textbf r,\textbf p,t) = f_0^{(\alpha)}(\textbf
r,\textbf p) + g^{(\alpha)}(\textbf r,\textbf p,t)$ ($\alpha=n$
for neutron and $\alpha=p$ for proton). Variations $g^{\alpha}$ of
the distribution function in external field are determined by the system
of the linearized Landau-Vlasov kinetic equations:
%\begin{equation}
%\label{land12} \left\{
%\begin{array}{cc}
%\frac{\partial g^p}{\partial t} + \frac{\textbf p}{m_p}\nabla g^p -
%  \nabla U_0^p(\textbf r) \nabla_{\textbf p} g^p & = \nabla \left(
%  \delta U^p(\textbf r, t) + \beta^pQ^p \right) \nabla_{\textbf p} f^p_0
%  + J^{p}(\textbf r,t), \\
%\frac{\partial g^n}{\partial t} + \frac{\textbf p}{m_n} \nabla g^n -
%  \nabla U_0^n(\textbf r)\nabla_{\textbf p} g^n & = \nabla \left(
%  \delta U^n(\textbf r, t) + \beta^nQ^n \right) \nabla_{\textbf p} f^n_0
%  + J^n(\textbf r,t),
%\end{array}
%\right.
%\end{equation}
\begin{eqnarray}
\label{land12}
  \frac{\partial g^{\alpha}}{\partial t} + \frac{\textbf p}{m_{\alpha}}
  \nabla g^{\alpha} - \nabla U_0^{\alpha}(\textbf r) \nabla_{\textbf p}
  g^{\alpha} - \nabla \left( \delta U^{\alpha}(\textbf r, t) +
  \beta^{\alpha}Q^{\alpha} \right) \nabla_{\textbf p} f^{\alpha}_0=
   J^{\alpha}
\end{eqnarray}
where $U^{\alpha}_0(\textbf r)$ is equilibrium component of the
self-consistence mean field and $\delta U^{\alpha}(\textbf r,
t)$ is deviation of the field. The external field $\beta(t)Q(r)$ has
the following form $V^{L}_{ext} = \beta(t)Q_{L}(r)Y_{L0}(\hat{r})$,
$Q_{L}(r) = r^L$, $\beta(t)=\beta_0\exp[-i(\omega+i\delta)t]$ with
$\beta_0\ll1$ and $\delta\to+0$.

Two equations of the system (\ref{land12}) are connected by the average
field:
\begin{eqnarray}
\label{land11}
  \delta U^{\alpha}(\textbf r, t) = \int d\textbf r' \,
  u^{\alpha \alpha}(\textbf
  r, \textbf r') \, \int d\textbf p' \, g^{\alpha}(\textbf r', \textbf p',
  t) +\nonumber \\ + \int d\textbf r' \, u^{\alpha \alpha'}(\textbf r,
  \textbf r')
  \, \int d\textbf p' \, g^{\alpha'}(\textbf r', \textbf p', t), \quad
  \alpha \neq \alpha'.
\end{eqnarray}

\section{Self-consistent strength function}

The Fourier transformation of the solution of the kinetic equation
system (\ref{land12}) can be written in the next general form
\begin{eqnarray}
\label{land50}
  g^{\alpha}(\textbf r,\textbf p, \omega) =
  g^{\alpha}_0(\textbf r,\textbf p,
  \omega) + \int d\textbf r' \int d\textbf p' \,
  \Omega^{\alpha}(\textbf r,\textbf r', \textbf p, \omega)
  g^{\alpha}(\textbf r',\textbf p', \omega) +\nonumber\\+
  \int d\textbf r' \int d\textbf p' \,
  \Omega^{\alpha'}(\textbf r,\textbf r',
  \textbf p, \omega) g^{\alpha'}(\textbf r',\textbf p', \omega),
\end{eqnarray}
where $g^{\alpha}_0(\textbf r,\textbf p, \omega)$ is
Fourier transformation of the solution \cite{Land1,torro}
of the equation (\ref{land12}) with
$\delta U^{\alpha}=0$, $J^{\alpha}=0$. The function
$\Omega^{\alpha}$ ($\Omega^{\alpha'}$) coincide with
$g^{\alpha}_0$ after replacing external field
$\beta^{\alpha}Q^{\alpha}$ by the interaction
$u^{\alpha\alpha}$ ($u^{\alpha\alpha'}$) and the solution
(\ref{land50}) can be rewritten as
\begin{eqnarray}
\label{landx11}
  g^{\alpha}(\textbf r,\textbf p, \omega) =
  \beta^{\alpha}(\omega)\int d\textbf r' \int d\textbf r'' \,
  \Omega^{\alpha}(\textbf r,\textbf r', \textbf p, \omega)
  D^{\alpha}(\textbf r',\textbf r'', \omega)Q^{\alpha}(\textbf r'') +\nonumber\\+
  g^{\alpha}_0(\textbf r,\textbf p,\omega) +
  \beta^{\alpha'}(\omega) \int d\textbf r' \int d\textbf r'' \,
  \Omega^{\alpha'}(\textbf r,\textbf r',\textbf p, \omega)
  D^{\alpha'}(\textbf r',\textbf r'', \omega)Q^{\alpha'}(\textbf
  r'')
\end{eqnarray}
with $D^{\alpha}(\textbf r,\textbf r',\omega)$ for the response
function
\begin{equation}
\label{land39}
  \delta n^{\alpha}(\textbf r, \omega) = \beta^{\alpha}(\omega)
  \int d\textbf r' \, D^{\alpha}(\textbf r,
  \textbf r', \omega)Q^{\alpha}(\textbf r') =
  \int d\textbf p' \, g^{\alpha}(\textbf r, \textbf p',\omega).
\end{equation}
Here, $\delta n^{\alpha}(\textbf r, \omega)$ is nucleon density
variation in external field. The system of integral equations for
response function $D^{\alpha}$ is obtained with the use of
(\ref{land39}) and integration (\ref{land11}) over momentum
$\textbf p$.
%\begin{eqnarray}
%\label{landx12}
%  \beta^{\alpha}(\omega)\int d\textbf r'\,D^{\alpha}_L
%  (\textbf r,\textbf r', \omega)Q^{\alpha}(\textbf r') =
%  \beta^{\alpha}(\omega)\int d\textbf r'\,D^{\alpha,0}_L
%  (\textbf r,\textbf r', \omega)Q^{\alpha}(\textbf r') +\nonumber\\+
%  \beta^{\alpha}(\omega)
%  \int d\textbf r'\int d\textbf r''\int d\textbf r'''\,
%  D^{\alpha,0}_L(\textbf r,\textbf r'', \omega)
%  u^{\alpha \alpha}(\textbf r'',\textbf r''')
%  D^{\alpha}_L(\textbf r''',\textbf r', \omega)
%  Q^{\alpha}(\textbf r')  +\nonumber\\+
%  \beta^{\alpha'}(\omega)
%  \int d\textbf r'\int d\textbf r''\int d\textbf r'''\,
%  D^{\alpha,0}_L(\textbf r,\textbf r'', \omega)
%  u^{\alpha \alpha'}(\textbf r'',\textbf r''')
%  D^{\alpha'}_L(\textbf r''',\textbf r', \omega)
%  Q^{\alpha'}(\textbf r'),
%\end{eqnarray}
%where $D^{\alpha,0}_L(\textbf r,\textbf r', \omega)$ --
%non-self-conjugated response function.

The strength function of nuclear response $S_L(\omega)$
determines the RS function and can be found in the following
form
\begin{equation}
\label{landx15a}
  S_L(\omega) = C^{(p)}_LS^{p}_L + C^{(n)}_LS^{n}_L,
\end{equation}
\begin{equation}
\label{land53}
   S^{\alpha}_L(\omega) \equiv -\frac{1}{\pi} Im
   \frac{\chi^{\alpha}+\chi^{\alpha}\chi^{\alpha'}
  (k^{\alpha\alpha'}\beta^{\alpha'}/
  \beta^{\alpha}-k^{\alpha\alpha})}
  {(1-k^{\alpha\alpha}\chi^{\alpha})(1-k^{\alpha'\alpha'}\chi^{\alpha'})
  -(k^{\alpha\alpha'})2\chi^{\alpha}\chi^{\alpha'}}.
\end{equation}
Here, $\chi^{\alpha}=\int d\textbf r \int d\textbf r' \,
Q_L(\textbf r)D_L^{\alpha,0}(\textbf r, \textbf r', \omega)
Q_L(\textbf  r')$.
The separable interaction potential $u^{\alpha \alpha'}_L(r, r')
=k^{\alpha \alpha'}(L)Q^{\alpha}(r) Q^{\alpha'}(r')$ was used.
The values of the coefficients $C^{(\alpha)}$ were found
from the energy weighted sum rule; $C^{(p)}_1=2N/A$,
$C^{(n)}_1=2Z/A$ for dipole electric external field.

The collision integral in equation (\ref{land12}) is taken in
relaxation times approximation with allowing for retardation
effects \cite{kolom,acta}, $J^{\alpha} = g^{\alpha}(\textbf r, \textbf
p,\omega) /\tau^{\alpha}(\omega)$. Here, $\tau(\omega)$ is
frequency and temperature dependent relaxation time. As a
result the $\chi^{\alpha}$ takes the following form
\cite{Land1,torro}
\begin{eqnarray}
\label{landx28}
  \chi^{\alpha} = \frac{8\pi^2}{2L+1}
  \sum_{n = -\infty}^{\infty} \sum_{N = -L}^L
  \left| Y_{LN}\left(\frac{\pi}2, \frac{\pi}2 \right) \right|^2
  \times \nonumber\\ \times
  \int d\epsilon \, \frac{\partial n_{\alpha}(\epsilon)}
  {\partial\epsilon}
  \int d\lambda \, \lambda \omega^{\alpha}_n(N) T^{\alpha}
  \frac{|Q^{\alpha}_{LM}(n, N)|^2}{\omega -
  \tilde\omega^{\alpha}(n,N)},
\end{eqnarray}
Where, $n_{\alpha}(\epsilon)$ is equilibrium distribution
function, $\omega^{\alpha}_n(N)$ -- eigenfrequencies \cite{torro}
in subsystem $\alpha$ without residual interaction,
$Q^{\alpha}_{LM}(n, N)$
-- matrix elements \cite{torro}.
Resonance frequency $\tilde\omega^{\alpha}(n,N)$
is obtained as a solution of the following equation:
\begin{equation}
\label{landx22}
  \omega -\frac{i}{\tau^{\alpha}(\omega)}=
  \omega_n^{\alpha}(N).
\end{equation}

\section{Calculations and conclusions}

%\begin{figure}[th]
%  \centerline{\includegraphics[width=0.7\textwidth,clip]{kfig1.eps}}
%  \caption{The RS function for $^{208}Pb$ calculated at different values
%           of the coupling constant $k^{np}$ \label{knp}}
%\end{figure}

The dipole photo-absorption RS function
$\overrightarrow{f}_{E1}(\epsilon_{\gamma}\equiv\hbar\omega)$
is connected with the strength function $S_{L=1}(\omega)$
and photo-absorption cross-section $\sigma_{E1}(\epsilon_{\gamma})$
in the following way \cite{Plujko2}:
\begin{equation}
  \overrightarrow{f}_{E1}(\epsilon_{\gamma}) \equiv
  \frac{\sigma _{E1}(\epsilon_{\gamma})}
  {3\epsilon_{\gamma}(\pi \hbar c)^2} =
  \frac{4\pi}9 \frac{e^{2}}{(\hbar c)^{3}}\cdot S_{L=1}(\omega),
  \; \epsilon_{\gamma}\equiv\hbar\omega.
\end{equation}

The photo-absorption strength functions for $^{208}Pb$ are shown
in Fig. 1. They were calculated with the
different values of the coupling constants $k^{np}$ and
infinite wall potential was taken for mean field.
The experiment data was taken from Ref. \cite{data}.
The Fermi-energies were calculated as in Fermi gas model.
The constants $k^{nn}$ and $k^{pp}$ were equal \cite{Bohr}
$k=k^{nn}=k^{pp}=113/A^{5/3}$.
The calculations show that the resonance peak is shifted to
higher energies with increasing in $k^{np}$ and
amplitude of the peak is decreased. The width of the
peak is increased.

The photo-absorption strength functions for the isobars with $A=208$
are shown in Fig. 2.
The resonance amplitude decreases and its width grows with asymmetry
coefficient $I=(N-Z)/A$. The energy of peak maximum is slowly 
dependent on $I$. The width of the RSF is proportional to the $I^2$.
An additional low-energy
peak appears at large asymmetry coefficient.

%\begin{figure}[th]
% \centerline{\includegraphics[width=0.7\textwidth,clip]{kfig2.eps}}
% \caption{The radiation strength function for isobars with $A=208$
% \label{isobar}}
%\end{figure}

The numerical analysis of RS function for asymmetric nuclei shows that the
RSF shape does not change strongly for the nuclei with $|I|\le0.25$
and due to this the one-component models of the RSF with modified
width could be used.

\newpage
\pagestyle{empty}

\begin{figure}
\begin{center}
\hspace{-10mm}
 \includegraphics[width=0.8\textwidth,clip]{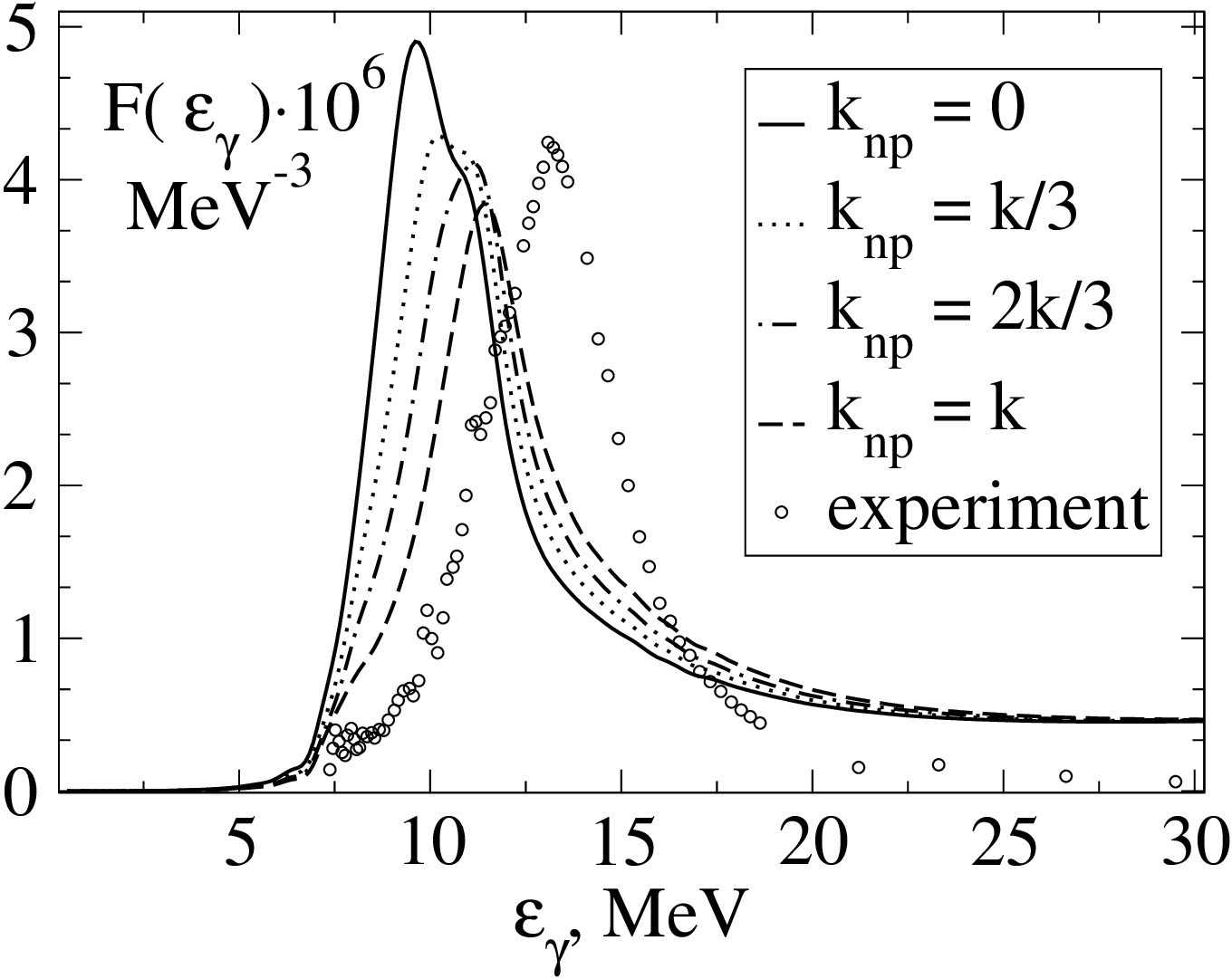}
\vspace{10mm}
 \parbox[t]{0.9\textwidth}{Fig.1: The RS function for
  $^{208}Pb$ calculated at different values of the
  coupling constant $k^{np}$}
 \end{center}
\end{figure}

\begin{figure}
\begin{center}
\hspace{-10mm}
 \includegraphics[width=0.8\textwidth,clip]{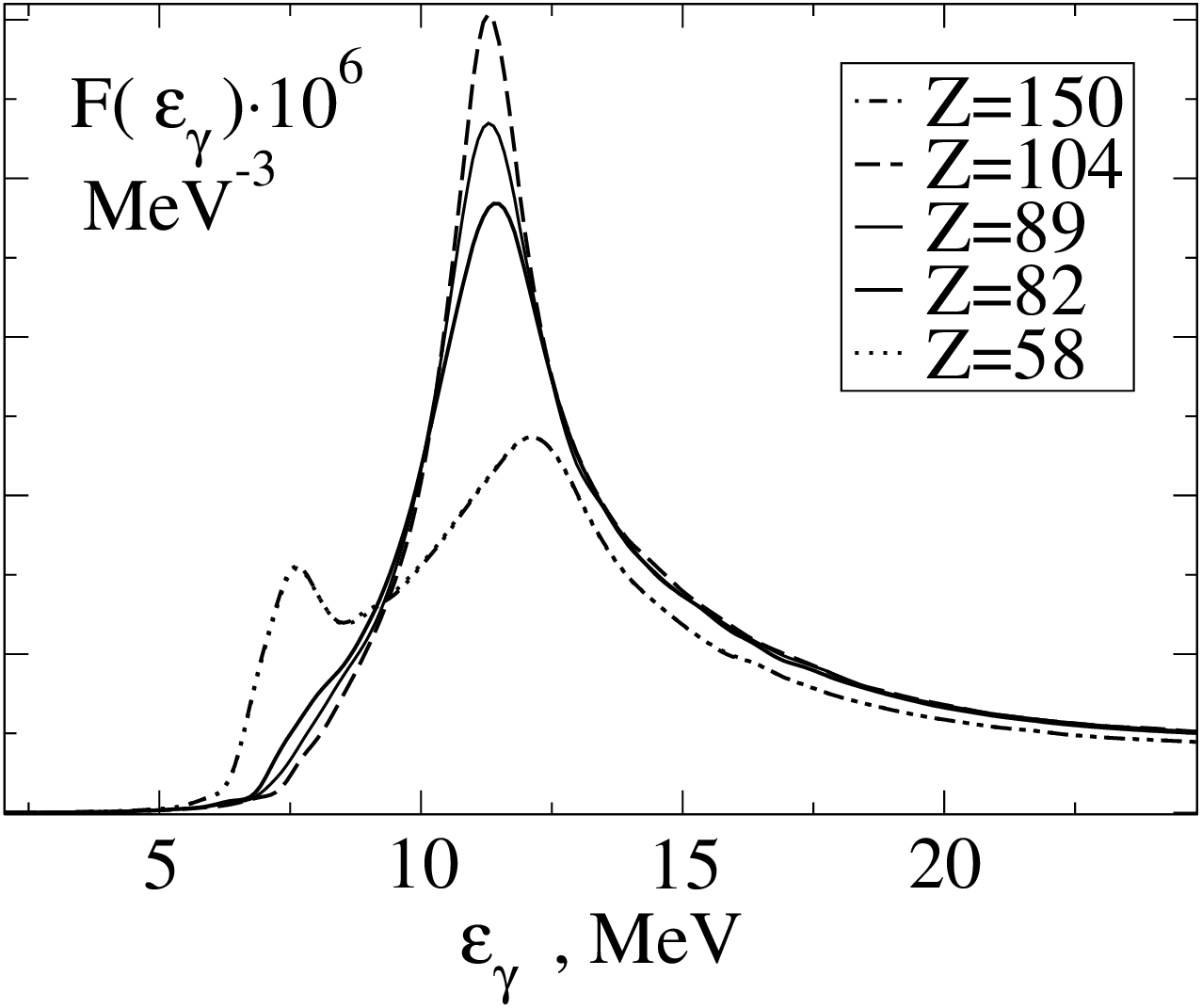}
\vspace{10mm}
 \parbox[t]{0.9\textwidth}{Fig.2: The radiation strength function
  for isobars with $A=208$}
 \end{center}
\end{figure}

\end{document}